\documentstyle[12pt,epsfig]{article}
\newcommand{\z}{&\hspace*{-8pt}}
\newcommand{\eps}{\varepsilon}

\begin{document}

\begin{center}
{\Large \bf  2-loop QCD corrections to $B_c$-meson leptonic
constant.}
\\ \vspace*{5mm} Andrei I. Onishchenko$^{a,b}$
and Oleg L. Veretin$^{a}$
\end{center}

\begin{center}
a) II. Institute f\"ur Theoretische Teilchenphysik,\\
Universit\"at Hamburg, 22761 Hamburg, Germany\\
\vspace*{0.5cm} 
b) Institute for High Energy Physics, Protvino, Russia \\
\end{center}

\vspace*{0.5cm}

\abstract{
We present results for two-loop QCD corrections to the short distance
coefficient that governs  leptonic decay of $B_c$-meson. After brief
discussion of the technics of asymptotic expansion
used to perform this calculation, we give
comments on the numerical value of two-loop corrections and its impact
on NRQCD description of $B_c$-meson bound state.
}

\section{Introduction}

The description of various aspects of $B_c$-meson physics%
\footnote{For a review see \cite{review}} has received recently a
lot of attention both from theoretical and experimental
sides. $B_c$-meson is the only meson within SM, which contains two
heavy quarks of different flavor. For this reason one might
expect, that its spectroscopy, production mechanisms and decays
differ significantly from those of charmonium, bottomonium as well
as hadrons with one heavy quark. Indeed, contrary to the case of
$J/\Psi$ and $\Upsilon$-mesons, $B_c$-meson is a long living
system, decaying trough electroweak interactions. To produce $\bar
b c$-system in flavor conserving collisions, an additional heavy
quark pair should be created, what makes the consideration of
production mechanism complex already at the leading order in the coupling
constant. The bound state heavy quarks dynamics, being very
similar to that of other quarkonia, at the same time has some
differences. With respect to the hierarchy of dynamical scales,
$B_c$-meson stands among the families of charmonium $\bar cc$ and
bottomonium $\bar bb$ and thus could be used to study both
quantitatively and conceptually existing effective low energy frameworks for
the description of bound state heavy quark dynamics, like NRQCD
\cite{NRQCD}, pNRQCD \cite{pNRQCD} and vNRQCD \cite{vNRQCD}. So, the
study of $B_c$-meson gives us an opportunity to test SM in a
completely new dynamical regime, explore it at a different angle,
what may give us an additional insight into our understanding of
SM in general.

Today, we already have many theoretical predictions, concerning
various $B_c$-meson properties: spectroscopy \cite{spect},
production \cite{prod} and decay channels \cite{decays}. In the
present work we will be interested in radiative corrections to
the leptonic decay constant of $B_c$-meson, entering expression for
its purely leptonic width. This particular decay mode of
$B_c$-meson is not really interesting for a direct measurement
--- the necessarily spin-flip makes its branching fraction very small.
However, the same quantity appears in the formulation of two-point
QCD sum rules \cite{QCDsr}. And the latter may be used to extract
an additional information about parameters of SM, namely the heavy
quark masses. A first step towards systematic formulation of NNLO
$B_c$-meson sum rules, treating $B_c$-meson as a Coulomb system,
was made in \cite{bcnlo}, where NLO analysis was presented. 
In order to describe $B_c$-meson QCD sum rules at NNLO, one needs both 2-loop
matching Wilson coefficient of QCD current with $B_c$ quantum
numbers on corresponding NRQCD current and two point NRQCD
correlator, computed at NNLO. Below we will present the results of
matching procedure. These results are also interesting from the
point of view of further development of computation technics used
in dimensionally regulated NRQCD \cite{Beneke1}. Here we are
dealing with threshold expansion with two mass scales $m_b$ and
$m_c$, what is an extension of previously obtained results for
matching of vector current with two equal heavy quark masses
\cite{Beneke2,Czarn}. The presence of two mass scales of course
makes problem more complicated, however as we will see the whole
calculation could be performed within the general framework of
threshold expansion in dimensionally regulated NRQCD \cite{Beneke1}. 
Here we would like to stress, that we treat $B_c$-meson as a Coulomb
system. So, in our matching procedure $c$-quark is considered to be heavy
with respect to factorization scale $m_c > \mu_{\rm fac}$. As a consequence, 
it is impossible to make connection of our result with the results on the
matching of heavy-light currents \cite{HL}. On the other side the limit
$m_b = m_c$ is well defined in our approach and in the case of vector current 
matching with equal quark masses we reproduce already known results of 
\cite{Beneke2,Czarn}.

This paper is organized as follows. In section 2 we introduce necessary
notation and present one-loop results obtained previously
\cite{Braaten}. Section 3 contains two-loop results for $B_c$-meson
leptonic decay constant. Here we also describe shortly various
steps of the performed calculation. And finally, in section 4 we
present our conclusion.


\section{$B_c$ - meson leptonic constant: one loop result}

The $B_c$-meson leptonic decay proceeds through a virtual $W$-boson.
The $W$-boson couples to the $B_c$ through the axial-vector part of
the charged weak current.  All QCD effects, both perturbative and
nonperturbative, enter into the decay rate through the decay
constant $f_{B_c}$, defined by the matrix element
\begin{equation}
\langle 0 | {\bar b} \gamma^\mu \gamma_5 c | B_c(P) \rangle \; =
\; i f_{B_c} P^\mu , \label{fbc}
\end{equation}
where $| B_c(P) \rangle$ is the state of $B_c$-meson with
four-momentum $P$.  It has the standard covariant normalization
$\langle B_c(P') | B_c(P) \rangle = (2 \pi)^3 2 E \,\delta^3(P' -
P)$ and its phase has been chosen so that $f_{B_c}$ is real and
positive.

To separate short-distance ($\sim 1/m_Q$) and long-distance ($\gg
1/m_Q $) QCD contributions to $f_{B_c}$ we use the formalism of
NRQCD \cite{NRQCD}. As a result the decay amplitude is factored
into short-distance coefficients multiplied by NRQCD matrix
elements. The Wilson short-distance coefficients could be then
calculated further order by order in perturbation theory by
matching a perturbative calculation in the full QCD with the
corresponding perturbative calculation in NRQCD. It is the subject
of this paper to calculate Wilson coefficient in front of NRQCD
current up to 2-loop order in perturbative QCD.

Let us now discuss the matching procedure in some detail. We start
with NRQCD lagrangian describing $B_c$-meson bound state dynamics
\begin{equation}
{\cal L}_{\rm NRQCD} \;=\; {\cal L}_{\rm light} \;+\;
\psi_c^\dagger \left(i D_0 + {\bf D}^2/(2 m_c) \right) \psi_c
\;+\; \chi_b^\dagger \left(i D_0 - {\bf D}^2/(2 m_b) \right)
\chi_b \;+\; \ldots , \label{NRQCD}
\end{equation}
where ${\cal L}_{\rm light}$ is the usual relativistic lagrangian
for gluons and light quarks.  The 2-component field $\psi_c$
annihilates charm quarks, while $\chi_b$ creates bottom
antiquarks. The typical relative velocity $v$ of heavy quarks
inside the meson provides a small parameter that can be used as a
nonperturbative expansion parameter.

To express the decay constant $f_{B_c}$ in terms of NRQCD matrix
elements, we  express the axial-vector current ${\bar b}
\gamma^\mu \gamma_5 c$ in terms of NRQCD fields.  Only the $\mu =
0$ component contributes to the matrix element (\ref{fbc}) in the
rest frame of the $B_c$-meson.  This component of the current has
the following operator expansion in terms of NRQCD fields:
\begin{equation}
{\bar b} \gamma^0 \gamma_5 c \;=\; C_0(m_b,m_c) \; \chi_b^\dagger
\psi_c \;+\; C_2(m_b,m_c) \; ({\bf D} \chi_b)^\dagger \cdot {\bf
D} \psi_c \;+\; \ldots, \label{opex}
\end{equation}
where $C_0$ and $C_2$ are Wilson short-distance coefficients that
depend on the quark masses $m_b$ and $m_c$.  By dimensional
analysis, the coefficient $C_2$ is proportional to $1/m_Q^2$.  The
contribution to the matrix element $\langle 0 | {\bar b} \gamma^0
\gamma_5 c | B_c \rangle$ from the operator $({\bf D}
\chi_b)^\dagger \cdot {\bf D} \psi_c$ is suppressed by $v^2$
relative to the operator $\chi_b^\dagger \psi_c$, where $v$ is the
relative velocity of heavy quarks inside $B_c$-meson. The dots in
(\ref{opex}) represent other operators whose contributions are
suppressed by higher powers of $v^2$.

The short-distance coefficient $C_0$ and $C_2$ can be determined
by matching perturbative calculations of the matrix elements in
the full QCD and NRQCD.  A convenient choice for matching is the
matrix element between the vacuum and the state $|c {\bar b}
\rangle$ consisting of a $c$ and a ${\bar b}$ on their perturbative
mass shells with nonrelativistic four-momenta $p$ and $p'$ in the
center of momentum frame:  ${\bf p} + {\bf p}' = 0$.  The matching
condition is
\begin{equation}
\langle 0 | {\bar b} \gamma^0 \gamma_5 c
    | c {\bar b} \rangle \Bigg|_{\mbox{{\scriptsize QCD}}}
= C_0 \; \langle 0 | \chi_b^\dagger \psi_c
    | c {\bar b} \rangle \Bigg|_{\mbox{{\scriptsize NRQCD}}}
 \;+\; C_2 \; \langle 0 | ({\bf D} \chi_b)^\dagger \cdot
 {\bf D} \psi_c | c {\bar b}\rangle \Bigg|_{\mbox{{\scriptsize NRQCD}}}
\;+\; \ldots, \label{match}
\end{equation}
where QCD and NRQCD represent perturbative QCD and perturbative
NRQCD, respectively. At leading order in $\alpha_s$, the matrix
element on the left hand side of (\ref{match}) is ${\bar v}_b(-{\bf p})
\gamma^0 \gamma_5 u_c({\bf p})$. The Dirac spinors are
\label{spinor}
\begin{eqnarray}
u_c(\mbox{{\bf p}}) &=& \sqrt{E_c+m_c \over 2E_c} \left(
\begin{array}{c}
    \xi \\
    {\mbox{{\bf p}} \cdot \mbox{{\boldmath $\sigma$}}
        \over E_c+m_c} \xi
\end{array}
    \right) ,
\label{uspinor}
\\
v(-\mbox{{\bf p}}) &=& \sqrt{E_b+m_b \over 2E_b}    \left(
\begin{array}{c}
    {(-\mbox{{\bf p}}) \cdot \mbox{{\boldmath $\sigma$}}
        \over E_b+m_b} \eta \\
    \eta
\end{array}
    \right) ,
\label{vspinor}
\end{eqnarray}
where $\xi$ and $\eta$ are 2-component spinors and $E_Q = m_Q^2 +
{\bf p}^2$.  Making a nonrelativistic expansion of the spinors to
second order in ${\bf p}/m_Q$, we find
\begin{equation}
{\bar v}_b(-{\bf p}) \gamma^0 \gamma_5 u_c({\bf p}) \;\approx\;
\eta_b^\dagger \xi_c \left( 1 \;-\; {1 \over 8} \left( {m_b + m_c
\over m_b m_c} \right)^2 {\bf p}^2
    \;+\; \ldots \right).
\end{equation}
At leading order in $\alpha_s$, the matrix elements on the right
hand side of (\ref{match}) are $\eta_b^\dagger \xi_c$ and ${\bf p}^2
\eta_b^\dagger \xi_c$. The short distance coefficients are
therefore $C_0 = 1$ and
\begin{equation}
C_2 \;=\; - {1 \over 8 \; m_{\rm red}^2} , \label{C2}
\end{equation}
where $m_{\rm red} = m_b m_c/(m_b+m_c)$ is the reduced mass of the
system.

In order to determine the short distance coefficients up to order
$\alpha_s^2$, we must calculate the matrix elements on both sides
of (\ref{match}) up to $\alpha_s^2$. Up to this order it is
sufficient to calculate radiative corrections only for the
coefficient $C_0$, since the contribution proportional to $C_2$ is
suppressed by additional power of $v^2$. The coefficient $C_0$ can be isolated by
taking the limit ${\bf p} \to 0$, in which case the matrix element
of $({\bf D} \chi_b)^\dagger \cdot {\bf D} \psi_c$ vanishes. Such
calculation at the one-loop order was performed in \cite{Braaten}, where
the following result was obtained for $C_0$:
\begin{equation}
C_0 \;=\; 1 \;+\; {\alpha_s(m_{\rm red}) \over \pi}
    \left[{m_b-m_c \over m_b+m_c} \log {m_b \over m_c}
        - 2 \right] .
\label{C0}
\end{equation}
$m_{\rm red}$ is the scale of the running coupling constant. To
the accuracy of this calculation, any scale of order $m_b$ or
$m_c$ would be equally correct.

Thus, the final result for the decay constant of the $B_c$ is
\begin{equation}
i f_{B_c} M_{B_c} \;=\; C_0 \; \langle 0 | \chi_b^\dagger \psi_c |
B_c \rangle \;+\; C_2 \; \langle 0 | ({\bf D} \chi_b)^\dagger
\cdot
    {\bf D} \psi_c | B_c \rangle .
\end{equation}
with short-distance coefficient $C_0$ up to next-to-leading order
in $\alpha_s$ as given in (\ref{C0}), while $C_2$ is given to
leading order in (\ref{C2}). The uncertainties consist of the
perturbative errors in the short distance coefficients and an
error of relative order $v^4$ from the neglected  matrix elements
that are higher order in $v^2$. The matrix element $\langle 0 |
\chi_b^\dagger \psi_c | B_c \rangle$ can be estimated using
wavefunctions at the origin from nonrelativistic potential models:
\begin{equation}
|\langle 0 | \chi_b^\dagger \psi_c | B_c \rangle|^2 \;\approx\; 2
M_{B_c} {3 \over 2 \pi} |R(0)|^2 .
\end{equation}
The factor of $2 M_{B_c}$ takes into account the relativistic
normalization of the state $|B_c \rangle$.


\section{Two-loop result}

In this section we consider the two-loop contribution to $B_c$-meson
leptonic constant. An analogous calculation, but for the case of
vector current and equal quark masses was done previously in
\cite{Beneke2,Czarn}. Here we are following the approach of
\cite{Beneke2,Czarn}, extending it, where it is necessary, to the
case of two different mass scales involved in our problem. In the
case of $B_c$-meson we consider the matching of QCD axial current
on the NRQCD one.

To compute Wilson short distance coefficient $C_0$ up to
$\alpha_s^2$ order, we replace $B_c$-meson state by a free
quark-antiquark pair of on-shell heavy quarks with small relative
velocity. In terms of the on-shell matrix elements for heavy
quarks with small relative velocity: $p = q m_1/(m_1 + m_2)$
and $p' = q m_2/(m_1 + m_2)$ ($q$ is the momentum of the gauge $W$-boson)
\footnote{This and previous definitions of external
heavy quark momenta lead to the same value of the matching
coefficient, but latter is more convenient in two-loop
calculations.}. The matching equation has the following form
\begin{equation}
 Z_{2,QCD}\Gamma_{QCD} =
C_0Z_{2,\rm NRQCD}Z_{J,\rm NRQCD}^{-1}\Gamma_{\rm NRQCD} + O(v^2), \label{matching}
\end{equation}
where $\Gamma_{\rm QCD}$ and $\Gamma_{\rm NRQCD}$ are amputated bare
axial-vector vertices in QCD and NRQCD, correspondingly. Here, we
would like to note, that since terms of order $v^2$ are not needed
to determine $C_0$ (see Eq. (\ref{match})), one can set  relative
momentum to zero and compute Green functions directly at
the threshold. Note, that here we deal with partially conserved
axial-vector current and its QCD renormalization constant is given
by $Z_{J,\rm QCD} = 1$, as in the case of vector current.

In order to obtain the coefficient $C_0$ at two loops one has to perform
the following steps:
\begin{itemize}
\item[1)] sum over all contributions to $\Gamma$'s, including the
one-loop term, multiplied by the two-loop QCD on-shell wave
function renormalization constant \cite{Z2onshell},
$Z_{J,\rm QCD}^{-1}$ and one-loop NRQCD-current renormalization
constant (at one-loop it is equal to unity), 
\item[2)] perform the one-loop renormalization of the coupling and masses.
\end{itemize}
After these manipulations the remaining poles in $\epsilon $
belong to the anomalous dimension of the corresponding NRQCD current and
finite part correspond to short-distance constant $C_0$.

In step 1) calculating two-loop contribution to vector-axial
vertex, it is convenient to write it in the following form
\begin{equation}
A_{\mu} = \bar b (p') \Bigl[ \gamma_{\mu}\gamma_5 F_1 (q^2) -
\frac{1}{m_1 + m_2}q_{\mu}\gamma_5 F_4 (q^2) \Bigr] c(p) \,.
\end{equation}
The formfactors $F_1 (q^2)$ and $F_2 (q^2)$ can be singled out by
imposing appropriate projections on them and thus our calculation
will reduce to the calculation of diagrams with nonvanishing
numerators.

\begin{figure}[t] 
\begin{center}
\leavevmode \epsfxsize=2cm \epsffile[220 410 420 540]{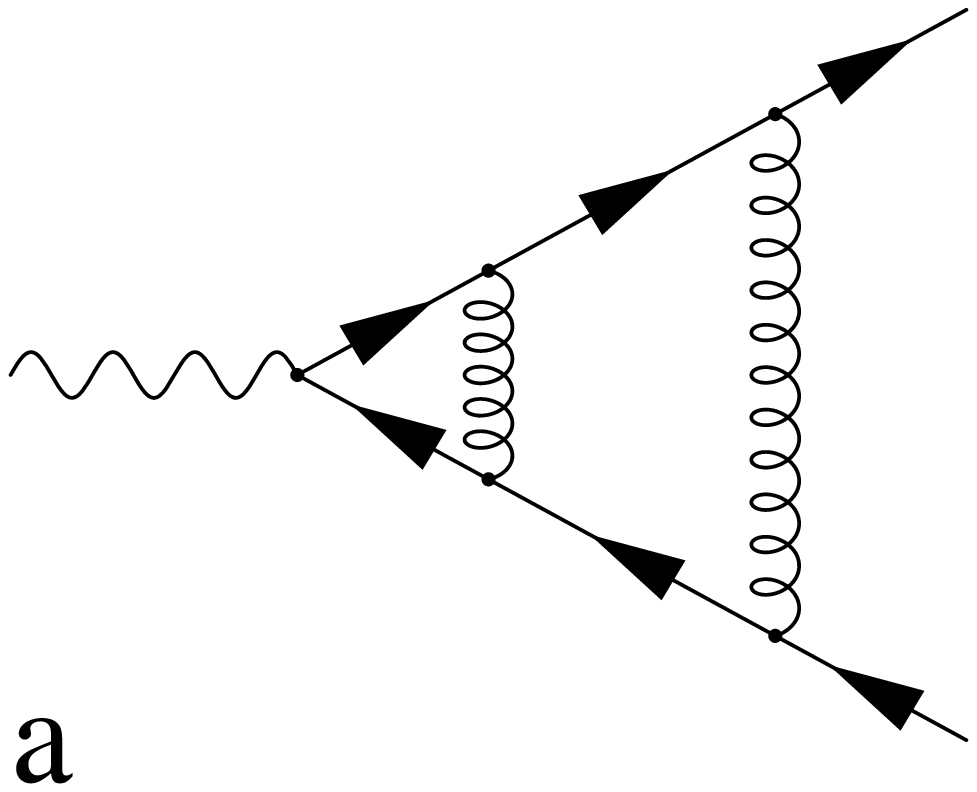}
\hspace{2cm} \leavevmode \epsfxsize=2cm \epsffile[220 410 420
540]{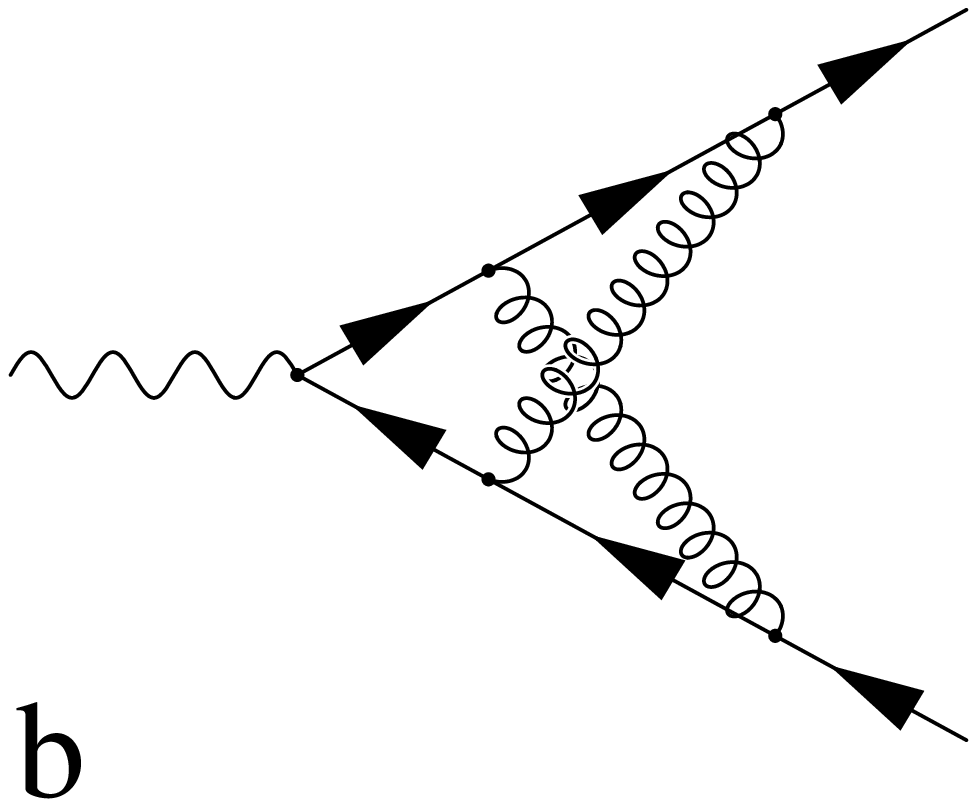} \hspace{2cm} \leavevmode \epsfxsize=2cm
\epsffile[220 410 420 540]{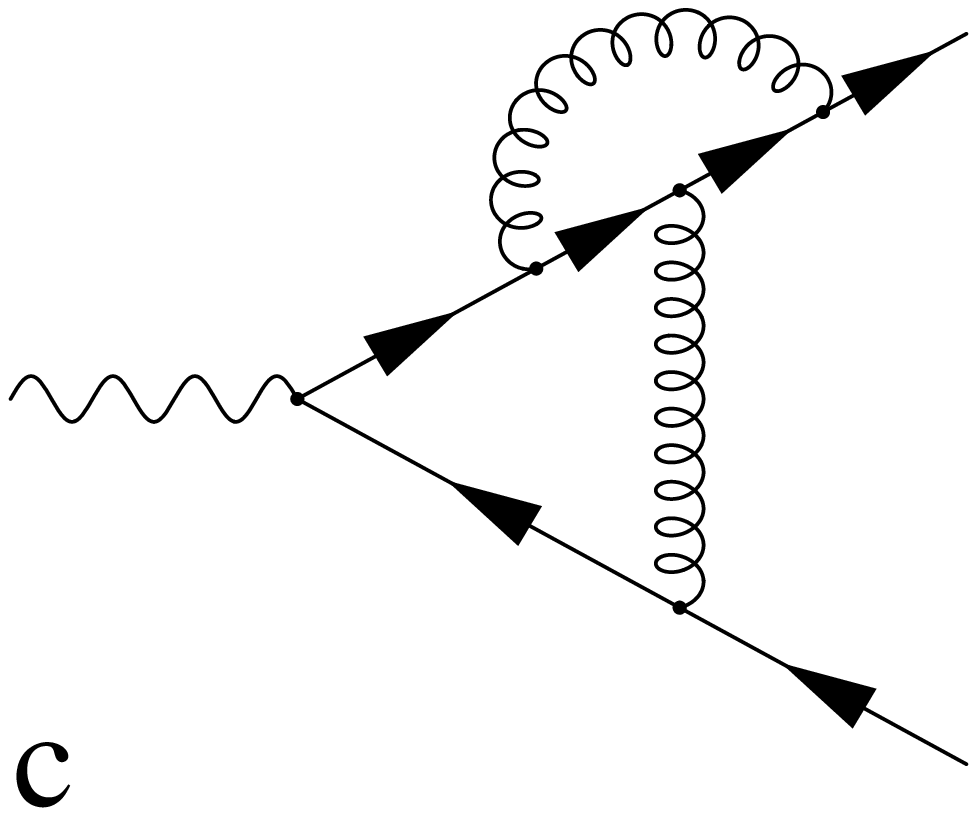} \hspace{2cm} \leavevmode
\epsfxsize=2cm \epsffile[220 410 420 540]{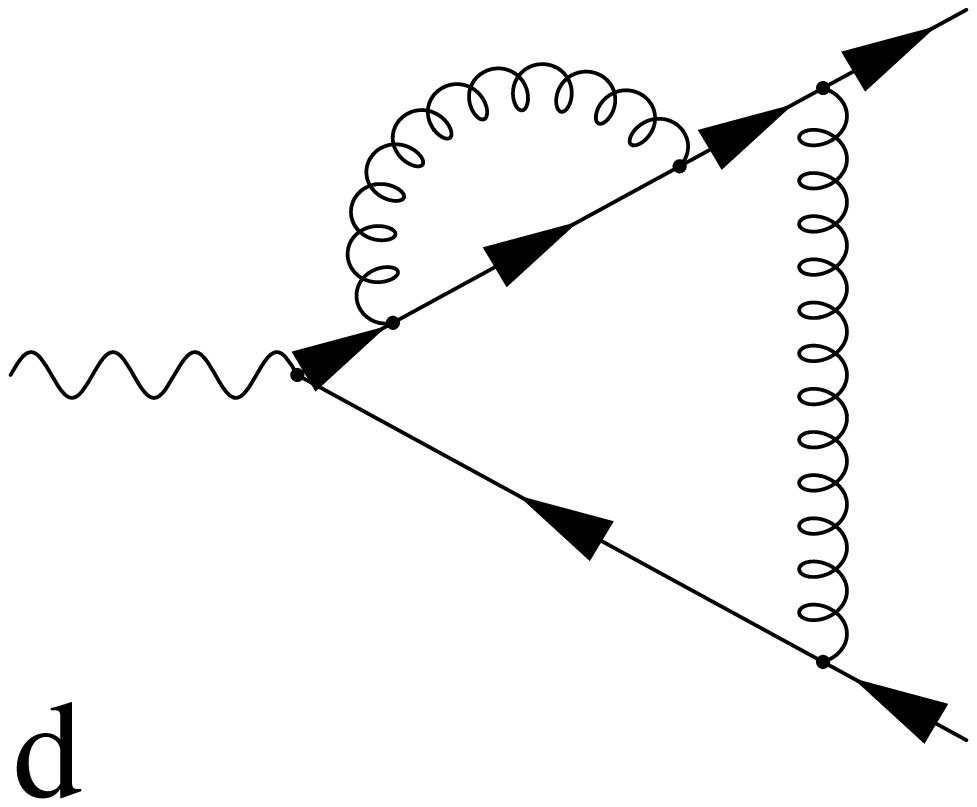}\\[1.5cm]
\leavevmode \epsfxsize=2cm \epsffile[220 410 420 540]{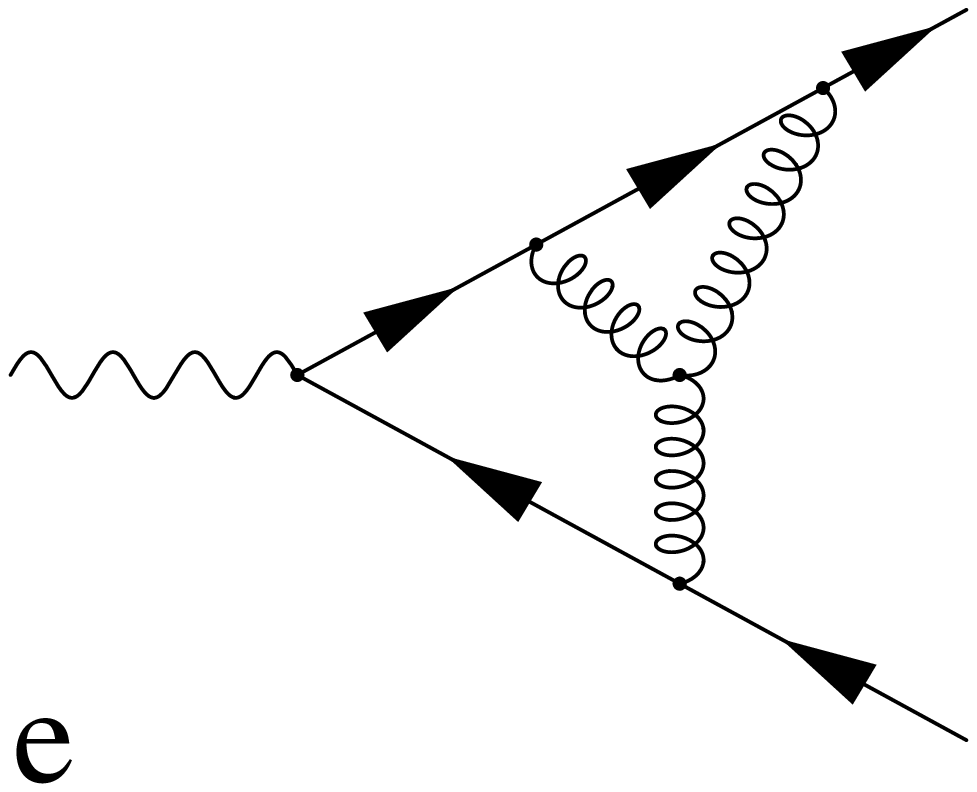}
\hspace{2cm} \leavevmode \epsfxsize=2cm \epsffile[220 410 420
540]{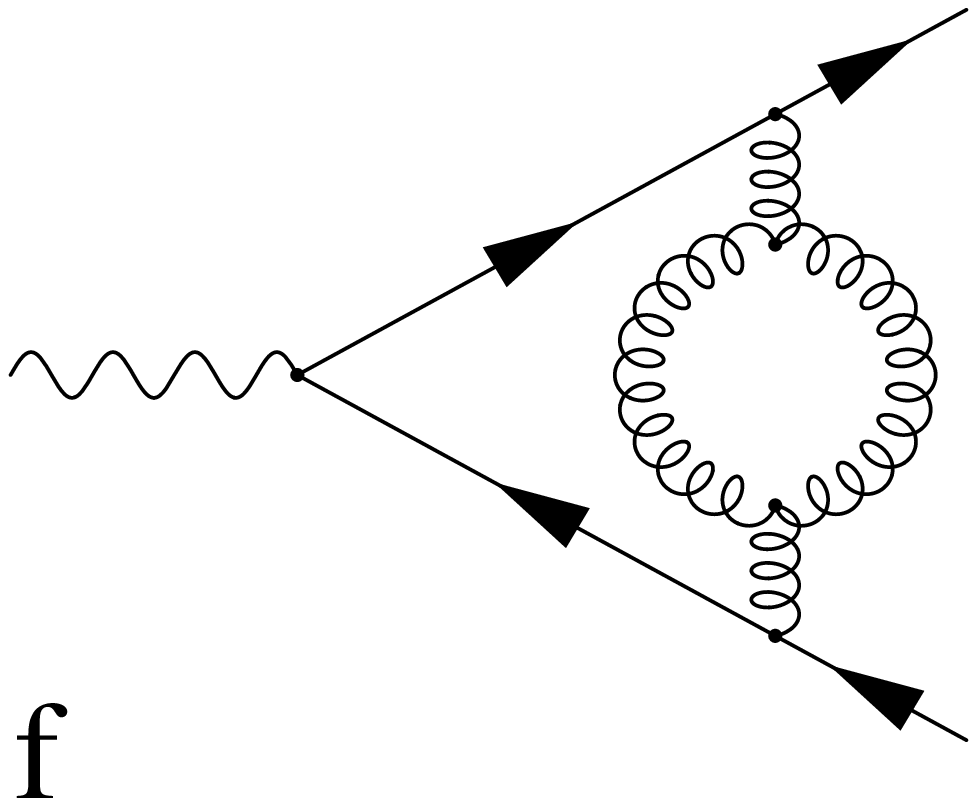} \hspace{2cm} \leavevmode \epsfxsize=2cm
\epsffile[220 410 420 540]{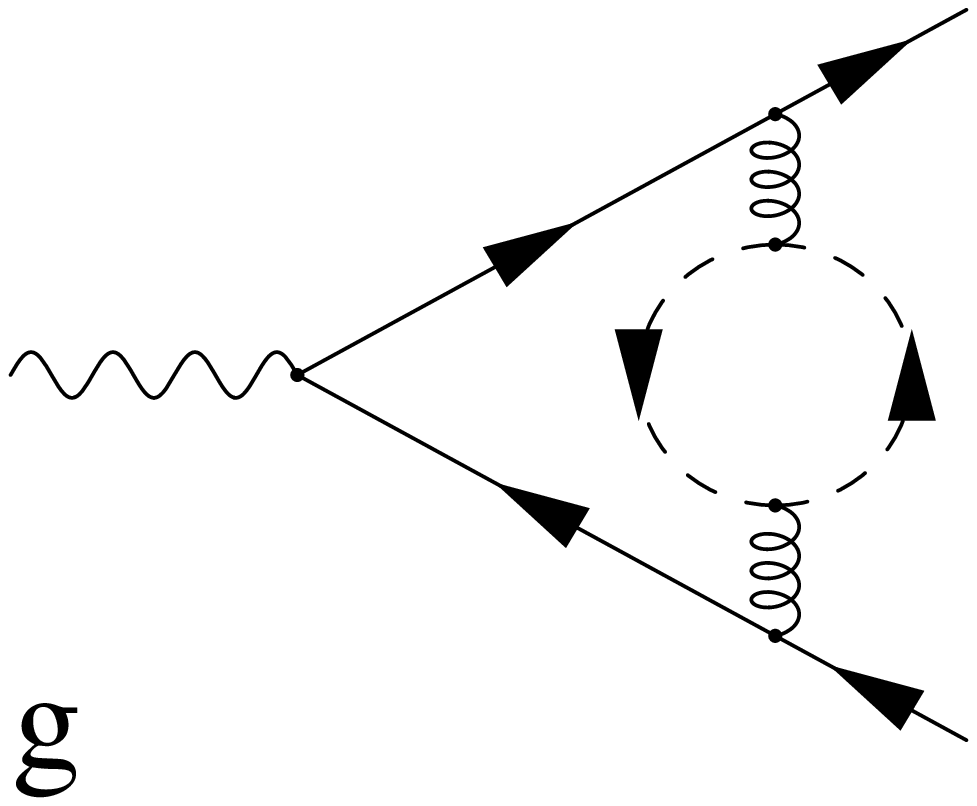} \hspace{2cm} \leavevmode
\epsfxsize=2cm \epsffile[220 410 420 540]{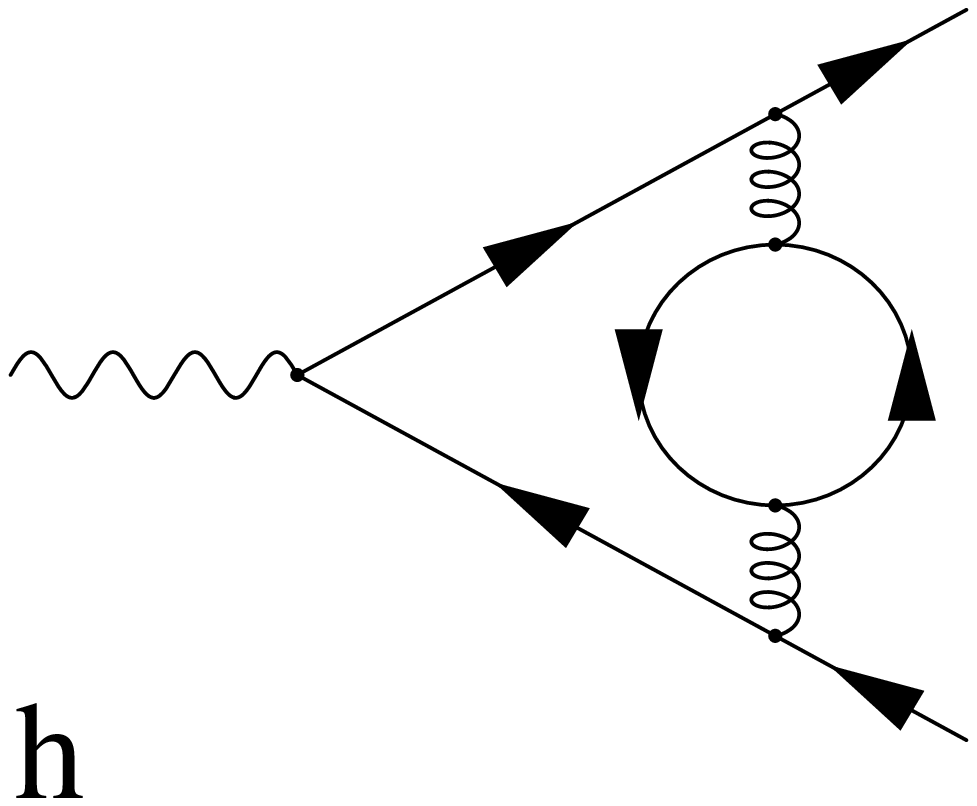}
\vskip  1.5cm
 \caption{\label{figQCDdiagrams}
QCD Feynman diagrams relevant for the calculation of the two-loop
corrections to the $B_c$ structure constant. }
\label{QCDdiagrams}
 \end{center}
\end{figure}

To simplify matching calculation at 2-loop order we use threshold
expansion in relative velocity $v$ of \cite{Beneke1}. The latter
is obtained by writing down contributions corresponding to hard
$(k\sim m_Q)$, soft $(k\sim m_Qv)$, potential $(k_0\sim m_Qv^2,
k_i\sim m_Qv)$ and ultrasoft $(k\sim m_Qv^2)$
regions\footnote{Here $k$ is the integration momentum.}. The
contributions from the last three regions can be easily identified
with NRQCD diagrams, that appear in the calculation of
$\Gamma_{\rm NRQCD}$. Thus, they simply drop out from the matching
relation (\ref{matching}) and it will suffice to compute only
contribution to threshold expansion of $\Gamma_{\rm QCD}$ of the
region, where all momenta are hard. Diagrams, whose hard
contribution need to be computed, are depicted in Fig.
\ref{QCDdiagrams}. Both form factors, introduced above have
Coulomb singularities at the threshold and diverge as $1/v^2$.
However, these singularities appear only in the soft, potential
and ultrasoft contributions, while the hard contribution is well
defined in dimensional regularization directly at the threshold. Once
the relative velocity is set to zero, the corresponding loop
integrals are considerably simplified and depend only on two
scales $M$ and $m$, where $M$ is the mass of the
heaviest quark and $m$ is the mass of the lighter quark. All loop integrals
with irreducible numerators can then be further reduced to the
scalar integrals with increased space-time dimension
\cite{Tarasovdim}. For the scalar integrals we derived recurrence
relations using integration by parts method of \cite{IBP}, which
allowed us to reduce all scalar integrals to sunsets. The sunset
integrals obtained were further reduced to master integrals with
the use of recurrence relations of
\cite{Tarasovgeneralized,DavydSmirn,DavydGrozin,Onishchenko:2002ri}. 
We would like to note, that our reduction procedure is valid for arbitrary values
of heavy quark masses, while to calculate some master integrals an
expansion in parameter $r=m/M$ was performed. 
All master integrals are of the sunset type discussed in the paper
\cite{Onishchenko:2002ri}. In particular there we discuss
in full details the procedure of expansion in parameter $r$.
Here, we describe only
shortly the calculation procedure and more detailed exposition
will be given elsewhere. The results, obtained for each 
of diagrams in Fig. \ref{QCDdiagrams} could be found in Appendix A.

After performing the steps of renormalization procedure described
above, we have the following $\overline{\mbox{MS}}$ expression for
the anomalous dimension of NRQCR current, which first arises here
at the two-loop order ($r=m/M$)
\begin{eqnarray}
\gamma_{J,\rm NRQCD} = \frac{d\ln Z_{J,\rm NRQCD}}{d\ln\mu}
= -\left\{
C_F^2\left(
2-\frac{(1-r)^2}{(1+r)^2}
\right)
+C_FC_A
\right\}\frac{\pi^2}{2}
\left(
\frac{\alpha_s}{\pi}
\right)^2+{\cal O}(\alpha_s^3).
\end{eqnarray}
This scale dependence is compensated by the scale dependence of the
two-loop matching coefficient $c_2(M/\mu)$, which is defined as
\begin{eqnarray}
C_0(\alpha_s,\frac{M}{\mu}) =
1+c_1(M/\mu)\frac{\alpha_s(M)}{\pi}
+c_2(M/\mu)
\left(
\frac{\alpha_s(M)}{\pi}
\right)^2
+\ldots,
\end{eqnarray}
In order to obtain $c_2$ in analytical form, we perform the expansion
of master integrals, considering the parameter $r=m/M$ as small
(in the practice $r=m_c/m_b\sim0.3$). Below we give the result of expansion
up to $O(r^2)$ inclusive.
Separating the different colour group factors, in $c_2(M/\mu)$ 
in the $\overline{\mbox{MS}}$ we have:
\begin{eqnarray}
c_2(M/\mu) \z=\z C_F^2c_{2,A}+C_FC_Ac_{2,NA}
+C_FT_Fn_lc_{2,L}+C_FT_Fc_{2,H}, \label{2loopcoeff} \\ 
\z\z \nonumber \\
c_{2,A} \z=\z
\left(
2-\frac{(1-r)^2}{(1+r)^2}
\right)
\frac{\pi^2}{4}\ln\left(\frac{M^2}{\mu^2}\right)
+\frac{9}{32}L^2+\frac{\pi^2}{3}L+\frac{41}{32}L
+\pi^2\ln (2)-\frac{91\pi^2}{48}+\frac{29}{16}
\nonumber \\
\z\z
+r\left[
\frac{17\pi^2}{12}L-\frac{43}{16}L-6\zeta_3
+\frac{23\pi^2}{24}-\frac{3}{8}
\right]
+r^2\left[
-\frac{21}{8}L^2-3\pi^2L+\frac{165}{16}L
+12\zeta_3-\frac{25\pi^2}{12}-\frac{147}{16}
\right],
\nonumber \\
\z\z \\
c_{2,NA} \z=\z
\frac{\pi^2}{4}\ln\left(\frac{M^2}{\mu^2}\right)
+\frac{11}{16}L^2+\frac{13\pi^2}{24}L+\frac{127}{96}L
-\frac{9}{4}\zeta_3-\frac{1}{2}\pi^2\ln (2)+\frac{41\pi^2}{96}
-\frac{23}{48}
\nonumber \\
\z\z
+r\left[
-\frac{25}{16}L^2-\frac{\pi^2}{24}L+\frac{65}{48}L-\frac{2\pi^2}{3}
-\frac{49}{32}
\right]
+r^2\left[
\frac{15}{8}L^2-\frac{23}{16}L+\frac{19\pi^2}{48}+\frac{31}{36}
\right], \nonumber \\
\z\z  \\
c_{2,L} \z=\z
-\frac{1}{4}L^2-\frac{11}{24}L+\frac{5\pi^2}{144}+\frac{1}{12}
+r\left[
\frac{L^2}{2}-\frac{L}{12}
\right]
+r^2\left[
\frac{L}{12}-\frac{L^2}{2}
\right], \\
c_{2,H} \z=\z
-\frac{L^2}{4}-\frac{11L}{24}-\frac{19\pi^2}{96}+\frac{1147}{288}
+r\left[
\frac{L^2}{2}-\frac{L}{12}-\frac{23\pi^2}{96}-\frac{103}{144}
\right]
\nonumber \\
\z\z
+r^2\left[
-\frac{L^2}{2}+\frac{89L}{60}+\frac{23\pi^2}{96}+\frac{11527}{3600}
\right].
\end{eqnarray}
Here $\zeta_3 = 1.202\ldots$, $M=m_b$, $r=m_c/m_b$,
$L\equiv\ln (r)$ and we have taken all fermions with masses less
than $m_b$ and $m_c$ as massless. The heavy quark masses, appearing 
in our formulae are pole quark masses. All calculations were performed
with the help of computer algebra system FORM \cite{Jos}. As a
check for our calculations, we employed a general covariant gauge.
To test our FORM programs we also have reproduced already known
results for the sort distance corrections to the vector current
with equal masses.

Now let us discuss the expression for 2-loop anomalous dimension
of $B_c$-meson NRQCD current. It is an exact formula in a sense,
that we resumed a whole series in $r$. In the limit of equal quark
masses it differs from the result obtained in the case of vector
current. This fact is explained by the contribution, coming from
Breit--Fermi potential, namely spin-spin and spin-orbital heavy
quark interaction, which in the case of $B_c$-meson has a
different expression compared to that computed for $J/\Psi$ and
$\Upsilon$-mesons \cite{Pivovarov}. This result for anomalous
expression could be also obtained considering NNLO corrections to
the Coulomb Green function, given by Breit--Fermi and non-abelian
potentials.

Numerically, the scale dependence of the two-loop matching coefficient 
(\ref{2loopcoeff}) is large. We expect it to be compensated
by the scale dependence of $B_c$-meson wave function computed within
the lattice NRQCD framework with the scale dependence of the results introduced
by imposing ultraviolet cutoff/factorization scale. The results
of nonrelativistic potential models should be considered as being
obtained at a scale typical for $B_c$-meson bound state dynamics
$\mu\sim 1$ GeV. For numerical estimates we consider $\mu = 1$ GeV
as an adequate bound state scale and use its value for factorization
scale $\mu_{\rm fac}$. In the case of $B_c$-meson and pole quark
masses ($m_b = 4.8$ GeV, $m_c = 1.65$ GeV) we have
\begin{eqnarray} 
f_{B_c} = \left[
1 - 1.48\left(\frac{\alpha_s(m_b)}{\pi}\right)
-24.24\left(\frac{\alpha_s(m_b)}{\pi}\right)^2
\right]f_{B_c}^{NR}
\end{eqnarray}
Here we see that 2-loop corrections are large and constitute $\approx 100\%$ of
1-loop correction. Even less favorable situation was detected earlier
for $J/\Psi$ states \cite{Beneke2}. The variation of factorization scale does not help to solve this
problem.  So, at this point one may conclude that application of NRQCD to $B_c$-meson
as well as for $J/\Psi$ is not well justified. However, while in the case of $J/\Psi$-meson
2-loop corrections far exceed 1-loop, one may speculate that situation in $B_c$-meson
is marginal, in a sense that one may still obtain sensible results using NRQCD framework
as well as treating $B_c$-meson as a heavy-light system. This way one may study 
the change in heavy quarks dynamics between bottomonium and charmonium families. Another
possible reason for such large correction is our use of particular renormalization
scheme $\overline{\mbox{MS}}$, which may turn out not to 
work well for heavy quark bound state 
systems. To prove or disprove this statement one has to calculate 2-loop corrections
to any relation between physical observables, from which 
quarkonium wave function at the origin 
is eliminated \cite{Beneke2}.

\section{Conclusion}
In this paper we presented the results for the two-loop
radiative corrections to the short distance coefficient in matching
between QCD and NRQCD for the $B_c$-meson currents. It turns out that the numerical value
of these corrections are large. There are several conclusions, which could be
made here. First NRQCD description is inadequate not only for charmonium
family but also for $B_c$-meson. To see whether it is true,  we suppose to perform a 
systematical study of NRQCD $B_c$-meson sum rules at NNLO in one of our 
next publications. Second, is it the use of 
$\overline{\mbox{MS}}$ renormalization scheme, which makes things look nasty?. To answer
this question one needs to compute some other decay mode of either $J/\Psi$ or $B_c$
-mesons  at 2-loop level and analyze perturbative corrections to the ratio of these
decay modes. 
  
We thank K.~Chetyrkin, A.~Davydychev, A.~Grozin, M.~Kalmykov, V.~Kiselev, 
O.~Tarasov and A.~Pivovarov 
for fruitful discussion of the topics discussed in this paper. We especially grateful
to V. Smirnov for encouragement, constant interest to this work and help with numerous 
checks of our calculation. This work was supported by DFG Forschergruppe
"Quantenfeldtheorie, Computeralgebra and Monte-Carlo-Simulation"
(contract FOR 264/2-1) and by BMBF under grant No 05HT9VKB0.


\appendix

\section{2 loop bare hard contribution }

Here we collected results for hard contributions of diagrams, shown in
Fig. \ref{QCDdiagrams}. Below, to save space, we present only
coefficient of
$\left(\alpha_s/\pi\right)^2(e^{\gamma_E}M^2/(4\pi\mu^2))^{-2\eps}$ for
each of the digrams, presented above.
Though our calculations were performed in general covariant gauge, 
we present here for brevity only the results in the Feynman gauge.
\begin{itemize}
\item {\bf diagram 1} (Fig. 1a)
\begin{eqnarray}
C_F^2\mbox{\Large \{}\z\z
\frac{9}{32\eps^2}+\frac{1}{\eps}\left(
-\frac{3}{64}-\frac{\pi^2}{4}-\frac{9L}{8}
+r[\frac{9L}{8}-\frac{\pi^2}{2}]
+r^2[\pi^2-\frac{9L}{8}]
 \right) \nonumber \\
\z\z+\frac{9L^2}{4}+\frac{11}{12}\pi^2L+\frac{L}{2}-3\zeta_3
-\frac{299\pi^2}{192}-\frac{41}{128}
+r\left[-3L^2+\frac{3}{2}\pi^2L-\frac{9L}{8}-6\zeta_3
+\frac{\pi^2}{3}-\frac{15}{32}\right] \nonumber \\
\z\z+r^2\left[\frac{7L^2}{8}-\frac{19}{6}\pi^2L+\frac{491L}{48}
+12\zeta_3-\frac{49\pi^2}{24}-\frac{401}{36}  \right]\mbox{\Large \}}
,
\end{eqnarray}
\item {\bf diagram 2} (Fig. 1d, with selfenergy for quark with mass $M$)
\begin{eqnarray}
C_F^2\mbox{\Large \{}\z\z
\frac{3}{32\eps^2}(1-\frac{2}{r})
+\frac{1}{\eps}\left(\frac{1}{r}[\frac{3L}{8}+\frac{3}{8}]
-\frac{3L}{8}-\frac{3}{64}-r[\frac{3L}{8}+\frac{3}{4}]
+r^2[\frac{9L}{8}+\frac{3}{4}] \right) \nonumber \\
\z\z-\frac{1}{r}\left[\frac{3}{8}L^2
+\frac{3}{4}L+\frac{\pi^2}{32}+\frac{13}{12} \right]
+\frac{9}{8}L^2+\frac{13}{8}L+\frac{45\pi^2}{64}+\frac{2041}{384}
\nonumber \\
\z\z+r\left[\frac{3}{8}L^2-\frac{17}{4}L+\frac{\pi^2}{4}-\frac{63}{32}\right]
+r^2\left[-\frac{5}{8}L^2+\frac{31}{6}L+\frac{\pi^2}{4}+\frac{1001}{288}
\right]\mbox{\Large \}},
\end{eqnarray}
\item {\bf diagram 3} (Fig. 1d, with selfenergy for quark with mass $m$)
\begin{eqnarray}
C_F^2\mbox{\Large \{}\z\z
\frac{3}{32\eps^2}(1-2r)+\frac{1}{\eps}\left(
-\frac{3}{8}L-\frac{51}{64}+r[\frac{3}{2}L+\frac{9}{8}]
-r^2[\frac{15}{8}L+\frac{3}{4}]\right) \nonumber \\
\z\z+\frac{3}{4}L^2+3L+\frac{247\pi^2}{192}+\frac{1057}{384}
-r\left[3L^2+\frac{29}{16}L+\frac{11\pi^2}{96}+\frac{37}{48}\right]
\nonumber \\ \z\z
+r^2\left[\frac{15}{4}L^2-\frac{11}{8}L-\frac{5\pi^2}{8}-\frac{21}{32} \right]
\mbox{\Large \}},
\end{eqnarray}
\item {\bf diagram 4} (Fig. 1e, with two gluons coupled to quark with mass $M$)
\begin{eqnarray}
C_FC_A\mbox{\Large \{}\z\z
\frac{15}{64\eps^2}+\frac{1}{\eps}\left(
\frac{19}{128}-\frac{9}{16}L+r[\frac{9}{16}L-\frac{\pi^2}{16}]
+r^2[\frac{\pi^2}{16}-\frac{9}{16}L]
\right) \nonumber \\
\z\z
+\frac{9}{16}L^2-\frac{3}{16}L-\frac{7\pi^2}{128}+\frac{329}{256}
+r\left[
-\frac{15}{16}L^2+\frac{\pi^2}{6}L+\frac{11}{16}L-\frac{3}{4}\zeta_3
-\frac{25\pi^2}{48}-\frac{21}{64}
\right]
\nonumber \\
\z\z
+r^2\left[
\frac{13}{16}L^2-\frac{\pi^2}{6}L-\frac{7}{96}L+\frac{3}{4}\zeta_3
+\frac{7\pi^2}{16}-\frac{115}{144}
\right]
\mbox{\Large \}},
\end{eqnarray}
\item {\bf diagram 5} (Fig. 1e, with two gluons coupled to quark with mass $m$)
\begin{eqnarray}
C_FC_A\mbox{\Large \{}\z\z
\frac{15}{64\eps^2}
+\frac{1}{\eps}\left(
-\frac{15}{16}L-\frac{\pi^2}{16}+\frac{19}{128}
+r[\frac{9}{16}L+\frac{\pi^2}{16}]
-r^2[\frac{9}{16}+\frac{\pi^2}{16} ]
\right) \nonumber \\
\z\z
+\frac{15}{8}L^2+\frac{\pi^2}{4}L-\frac{5}{16}L-\frac{3}{4}\zeta_3
-\frac{221\pi^2}{384}+\frac{473}{256}
+r\left[
-\frac{3}{2}L^2-\frac{\pi^2}{4}L+\frac{3}{16}L+\frac{3}{4}\zeta_3
+\frac{\pi^2}{12}-\frac{21}{64}
\right]
\nonumber \\
\z\z
+r^2\left[
2L^2+\frac{\pi^2}{4}L-\frac{83}{96}L-\frac{3}{4}\zeta_3+\frac{\pi^2}{8}
+\frac{353}{288}
\right]
\mbox{\Large \}},
\end{eqnarray}
\item {\bf diagram 6} (Fig. 1c, with two gluons coupled to quark with mass $M$)
\begin{eqnarray}
C_F(C_A-2C_F)\mbox{\Large \{}\z\z
-\frac{9}{64\eps^2}
+\frac{1}{\eps}\left(
\frac{3}{16}L-\frac{27}{128}-\frac{3}{16}rL+\frac{3}{16}r^2L
\right)
+\frac{3}{16}L^2+\frac{5}{8}L-\frac{3}{16}\zeta_3
\nonumber \\
\z\z
+\frac{\pi^2}{8}\ln (2)+\frac{33\pi^2}{128}+\frac{287}{256}
+r\left[
\frac{3}{16}L^2-L-\frac{3}{8}\zeta_3+\frac{\pi^2}{4}\ln (2)
+\frac{\pi^2}{12}-\frac{9}{64}
\right]
\nonumber \\
\z\z
+r^2\left[
\frac{1}{16}L^2-\frac{11}{48}L+\frac{3}{8}\zeta_3
-\frac{\pi^2}{4}\ln (2)+\frac{5\pi^2}{48}+\frac{71}{36}
\right]
\mbox{\Large \}},
\end{eqnarray}
\item {\bf diagram 7} (Fig. 1c, with two gluons coupled to quark with mass $m$)
\begin{eqnarray}
C_F(C_A-2C_F)\mbox{\Large \{}\z\z
-\frac{9}{64\eps^2}
+\frac{1}{\eps}\left(
\frac{9}{16}L-\frac{27}{128}-\frac{3}{16}rL+\frac{3}{16}r^2L
\right)
-\frac{9}{8}L^2+\frac{3}{4}L-\frac{3}{16}\zeta_3
\nonumber \\
\z\z
+\frac{5}{8}\pi^2\ln (2)+\frac{77\pi^2}{128}+\frac{215}{256}
+r\left[
\frac{3}{8}L^2+\frac{25}{32}L-\frac{3}{8}\zeta_3-\frac{3}{4}\pi^2\ln (2)
-\frac{9\pi^2}{32}-\frac{53}{32}
\right]
\nonumber \\
\z\z
+r^2\left[
-\frac{3}{16}L^2-\frac{5}{4}L+\frac{3}{8}\zeta_3+\frac{3}{4}\pi^2\ln (2)
+\frac{11\pi^2}{24}+\frac{137}{64}
\right]
\mbox{\Large \}},
\end{eqnarray}
\item {\bf diagram 8} (Fig. 1b)
\begin{eqnarray}
C_F(C_A-2C_F)\mbox{\Large \{}\z\z
\frac{1}{\eps}\left(
\frac{3}{16}-\frac{\pi^2}{16}
\right)
+\frac{7}{24}\pi^2L-\frac{5}{8}L-\frac{15}{8}\zeta_3-\frac{3}{4}\pi^2\ln (2)
-\frac{3\pi^2}{16}-\frac{35}{12} \nonumber \\
\z\z
+r\left[
\frac{3}{16}L^2+\frac{\pi^2}{24}L-\frac{5}{32}L+\frac{3}{4}\zeta_3
+\frac{\pi^2}{2}\ln (2)-\frac{\pi^2}{32}+\frac{59}{64}
\right]
\nonumber \\
\z\z
+r^2\left[
-\frac{11}{16}L^2-\frac{\pi^2}{12}L+\frac{11}{6}L-\frac{3}{4}\zeta_3
-\frac{\pi^2}{2}\ln (2)-\frac{35\pi^2}{48}-\frac{2119}{576}
\right]
\mbox{\Large \}},\quad\quad
\end{eqnarray}
\item {\bf diagram 9} (Fig. 1f)
\begin{eqnarray}
C_FC_A\mbox{\Large \{}\z\z
\frac{19}{128\eps^2}
+\frac{1}{\eps}\left(
\frac{33}{256}-\frac{19}{32}L+\frac{19}{32}rL-\frac{19}{32}r^2L
\right)
+\frac{19}{16}L^2-\frac{7}{32}L+\frac{95\pi^2}{768}+\frac{1315}{512}
\nonumber \\
\z\z
-r\left[
\frac{19}{16}L^2+\frac{5}{64}L
\right]
+r^2\left[
\frac{19}{16}L^2+\frac{5}{64}L
\right]
\mbox{\Large \}},
\end{eqnarray}
\item {\bf diagram 10} (Fig. 1g)
\begin{eqnarray}
C_FC_A\mbox{\Large \{}\z\z
\frac{1}{128\eps^2}
+\frac{1}{\eps}\left(
\frac{3}{256}-\frac{1}{32}L+\frac{1}{32}rL-\frac{1}{32}r^2L
\right)
+\frac{1}{16}L^2-\frac{1}{32}L+\frac{5\pi^2}{768}+\frac{73}{512}
\nonumber \\
\z\z
+r\left[
\frac{1}{64}L-\frac{1}{16}L^2
\right]
+r^2\left[
\frac{1}{16}L^2-\frac{1}{64}L
\right]
\mbox{\Large \}},
\end{eqnarray}
\item {\bf diagram 11} (Fig. 1h, with light quark loop)
\begin{eqnarray}
C_FT_Fn_l\mbox{\Large \{}\z\z
-\frac{1}{8\eps^2}
+\frac{1}{\eps}\left(
\frac{1}{2}L-\frac{1}{16}-\frac{1}{2}r(1-r)L
\right)
-L^2-\frac{5\pi^2}{48}-\frac{67}{32}
+r(1-r)\left[
L^2+\frac{1}{4}L
\right]
\mbox{\Large \}}, \nonumber \\
\z\z
\end{eqnarray}
\item {\bf diagram 12} (Fig. 1h, with quark of mass $M$ in the loop)
\begin{eqnarray}
C_FT_F\mbox{\Large \{}\z\z
-\frac{1}{4\eps^2}
+\frac{1}{\eps}\left(
\frac{1}{2}L+\frac{5}{48}-\frac{1}{2}r(1-r)L
\right)
-\frac{1}{2}L^2-\frac{1}{3}L-\frac{\pi^2}{8}-\frac{3}{32}
\nonumber \\
\z\z
+r\left[
\frac{1}{2}L^2+\frac{1}{3}L+\frac{\pi^2}{6}-\frac{211}{144}
\right]
+r^2\left[
-\frac{1}{2}L^2-\frac{13}{30}L-\frac{\pi^2}{6}+\frac{6007}{3600}
\right]
\mbox{\Large \}},
\end{eqnarray}
\item {\bf diagram 13} (Fig. 1h, with quark of mass $m$ in the loop)
\begin{eqnarray}
C_FT_F\mbox{\Large \{}\z\z
-\frac{1}{4\eps^2}
+\frac{1}{\eps}\left(
L+\frac{5}{48}-\frac{1}{2}r(1-r)
\right)
-2L^2-\frac{2}{3}L+\frac{13\pi^2}{96}-\frac{563}{288}
\nonumber \\
\z\z
+r\left[
L^2+\frac{1}{4}L-\frac{7\pi^2}{32}+\frac{3}{4}
\right]
+r^2\left[
-L^2+\frac{5}{4}L+\frac{13\pi^2}{32}
\right]
\mbox{\Large \}},
\end{eqnarray}
\end{itemize}
where $L\equiv \ln r$ and $n_l$ is the number of light fermions.


\begin{thebibliography}{**}

\bibitem{review}
  I.~P.~Gouz, V.~V.~Kiselev, A.~K.~Likhoded, V.~I.~Romanovsky and O.~P.~Yushchenko,
  Phys.\ Atom.\ Nucl.\  {\bf 67} (2004) 1559
  [Yad.\ Fiz.\  {\bf 67} (2004) 1581];\\
V.~V.~Kiselev,
[hep-ph/0211021]; 
%
S.~S.~Gershtein {\it et al.},
[hep-ph/9803433]; \\
S.~S.~Gershtein, V.~V.~Kiselev, A.~K.~Likhoded and
A.~V.~Tkabladze,
Phys.\ Usp.\  {\bf 38} (1995) 1 ; [Usp.\ Fiz.\ Nauk {\bf 165}
(1995) 3].

\bibitem{NRQCD}
G.T.~Bodwin, E.~Braaten and  G.P.Lepage, Phys. Rev. D {\bf 51}
(1995) 1125; [Erratum-ibid. {\bf 55} (1995) 5853];\\
T.~Mannel, G.A.~Schuler, Z. Phys. C {\bf 67} (1995) 159.

\bibitem{pNRQCD}
N.~Brambilla, A.~Pineda, J.~Soto and A.~Vairo,
Nucl.\ Phys.\ B {\bf 566} (2000) 275; \\
N.~Brambilla, A.~Pineda, J.~Soto and A.~Vairo,
Phys.\ Rev.\ D {\bf 60} (1999) 091502.

\bibitem{vNRQCD}
A.~V.~Manohar and I.~W.~Stewart,
Phys.\ Rev.\ D {\bf 62} (2000) 074015;\\
A.~V.~Manohar and I.~W.~Stewart,
Phys.\ Rev.\ D {\bf 63} (2001) 054004.

\bibitem{spect}
E.~J.~Eichten and C.~Quigg,
Phys.\ Rev.\ D {\bf 49} (1994) 5845; \\
V.~V.~Kiselev, A.~K.~Likhoded and A.~V.~Tkabladze,
Phys.\ Rev.\ D {\bf 51} (1995) 3613.

\bibitem{prod}
C.~H.~Chang and Y.~Q.~Chen,
Phys.\ Rev.\ D {\bf 46} (1992) 3845 [Erratum-ibid.\ D {\bf 50} (1994) 6013]; \\
E.~Braaten, K.~m.~Cheung and T.~C.~Yuan,
Phys.\ Rev.\ D {\bf 48} (1993) 5049; \\
V.~V.~Kiselev, A.~K.~Likhoded and M.~V.~Shevlyagin,
Z.\ Phys.\ C {\bf 63} (1994) 77; \\
K.~Kolodziej, A.~Leike and R.~Ruckl,
Phys.\ Lett.\ B {\bf 348} (1995) 219; \\
K.~Kolodziej, A.~Leike and R.~Ruckl,
Phys.\ Lett.\ B {\bf 355} (1995) 337; \\
A.~V.~Berezhnoi, V.~V.~Kiselev and A.~K.~Likhoded,
Phys.\ Lett.\ B {\bf 381} (1996) 341; \\
A.~V.~Berezhnoi, V.~V.~Kiselev and A.~K.~Likhoded,
Z.\ Phys.\ A {\bf 356} (1996) 79.

\bibitem{decays}
M.~Lusignoli, M.~Masetti, Z. Phys. C {\bf 51} (1991) 549; \\
V.~V.~Kiselev and A.~V.~Tkabladze,
Sov.\ J.\ Nucl.\ Phys.\  {\bf 48} (1988) 341 [Yad.\ Fiz.\  {\bf
48} (1988) 536]; \\
V.~V.~Kiselev and A.~V.~Tkabladze,
Phys.\ Rev.\ D {\bf 48} (1993) 5208; \\
C.~H.~Chang and Y.~Q.~Chen,
Phys.\ Rev.\ D {\bf 49} (1994) 3399; \\
M.~A.~Ivanov, J.~G.~Korner and P.~Santorelli,
Phys.\ Rev.\ D {\bf 63} (2001) 074010; \\
D.~Scora and N.~Isgur,
Phys.\ Rev.\ D {\bf 52} (1995) 2783; \\
A.~Y.~Anisimov, P.~Y.~Kulikov, I.~M.~Narodetsky and
K.~A.~Ter-Martirosian,
Phys.\ Atom.\ Nucl.\  {\bf 62} (1999) 1739 [Yad.\ Fiz.\  {\bf
62} (1999) 1868]; \\
I.Bigi, Phys. Lett. {\bf B371} (1996) 105;\\
M.Beneke, G.Buchalla, {Phys. Rev.} D {\bf 53} (1996) 4991;\\
G.~Chiladze, A.~F.~Falk and A.~A.~Petrov,
Phys.\ Rev.\ D {\bf 60} (1999) 034011;\\
A.~I.~Onishchenko, [hep-ph/9912424];\\
V.~V.~Kiselev, A.~K.~Likhoded, A.~I.~Onishchenko, Nucl. Phys. B {\bf 569} (2000) 473; \\
V.~V.~Kiselev, A.~K.~Likhoded, A.~E.~Kovalsky, Nucl. Phys. B {\bf 585} (2000) 353; \\
T.~Mannel and S.~Wolf,
Phys.\ Rev.\ D {\bf 65} (2002) 074012.

\bibitem{QCDsr}
M.~A.~Shifman, A.~I.~Vainshtein and V.~I.~Zakharov,
Nucl.\ Phys.\ B {\bf 147} (1979) 448,
Nucl.\ Phys.\ B {\bf 147} (1979) 385,
Nucl.\ Phys.\ B {\bf 147} (1979) 519;\\
V.~A.~Novikov, L.~B.~Okun, M.~A.~Shifman, A.~I.~Vainshtein,
M.~B.~Voloshin and V.~I.~Zakharov,
Phys.\ Rept.\  {\bf 41} (1978) 1;\\
L.~J.~Reinders, H.~Rubinstein and S.~Yazaki,
Phys.\ Rept.\  {\bf 127} (1985) 1.

\bibitem{bcnlo}
A.~I.~Onishchenko,
[hep-ph/0005127].

\bibitem{Beneke1}
M.~Beneke and V.~A.~Smirnov,
Nucl.\ Phys.\ B {\bf 522} (1998) 321; \\
V.~A.~Smirnov,
{\it  Berlin, Germany: Springer (2002) 262 p}; \\
V.~A.~Smirnov,
in {\it Proc. of the 5th International Symposium on Radiative
Corrections (RADCOR 2000) } ed. Howard E. Haber, [hep-ph/0101152];
\\
V.~A.~Smirnov,
Phys.\ Lett.\ B {\bf 465} (1999) 226; \\
M.~Beneke, A.~Signer and V.~A.~Smirnov,
[hep-ph/9906476]; \\
V.~A.~Smirnov and E.~R.~Rakhmetov,
Theor.\ Math.\ Phys.\  {\bf 120} (1999) 870 [Teor.\ Mat.\ Fiz.\
{\bf 120} (1999) 64]; \\
V.~A.~Smirnov,
[hep-ph/9708423]; \\
A.~Czarnecki and V.~A.~Smirnov,
Phys.\ Lett.\ B {\bf 394} (1997) 211; \\
V.~A.~Smirnov,
Phys.\ Lett.\ B {\bf 394} (1997) 205.

\bibitem{Beneke2}
M.~Beneke, A.~Signer and V.~A.~Smirnov,
Phys.\ Rev.\ Lett.\  {\bf 80} (1998) 2535.

\bibitem{Czarn}
A.~Czarnecki and K.~Melnikov,
Phys.\ Rev.\ Lett.\  {\bf 80} (1998) 2531.

\bibitem{HL}
D.~J.~Broadhurst and A.~G.~Grozin,
Phys.\ Rev.\ D {\bf 52} (1995) 4082; \\
A.~Czarnecki and K.~Melnikov,
Phys.\ Rev.\ D {\bf 66} (2002) 011502.

\bibitem{Braaten}
E.~Braaten and S.~Fleming,
Phys.\ Rev.\ D {\bf 52} (1995) 181.

\bibitem{Z2onshell}
N.~Gray, D.~J.~Broadhurst, W.~Grafe and K.~Schilcher,
Z.\ Phys.\ C {\bf 48} (1990) 673; \\
D.~J.~Broadhurst, N.~Gray and K.~Schilcher,
Z.\ Phys.\ C {\bf 52} (1991) 111; \\
J.~Fleischer, F.~Jegerlehner, O.~V.~Tarasov and O.~L.~Veretin,
Nucl.\ Phys.\ B {\bf 539} (1999) 671 [Erratum-ibid.\ B {\bf 571} (2000) 511; \\
K.~Melnikov and T.~van Ritbergen,
Nucl.\ Phys.\ B {\bf 591} (2000) 515.

\bibitem{Tarasovdim}
O.~V.~Tarasov,
Nucl.\ Phys.\ Proc.\ Suppl.\  {\bf 89} (2000) 237; \\
O.~V.~Tarasov,
Phys.\ Rev.\ D {\bf 54} (1996) 6479.

\bibitem{IBP}
F.V.~Tkachov, Phys. Lett. B {\bf 100} (1981) 65; \\
K.G.~Chetyrkin and F.V.~Tkachov, Nucl. Phys. B {\bf 192} (1981) 159.

\bibitem{Tarasovgeneralized}
O.V.~Tarasov, Nucl. Phys. B {\bf 502} (1997) 455.

\bibitem{DavydSmirn}
A.I.~Davydychev and V.A.~Smirnov, Nucl. Phys. B {\bf 554} (1999) 391.

\bibitem{DavydGrozin}
A.~I.~Davydychev and A.~G.~Grozin,
Phys.\ Rev.\ D {\bf 59} (1999) 054023.

\bibitem{Onishchenko:2002ri}
  A.~Onishchenko and O.~Veretin,
  Phys.\ Atom.\ Nucl.\  {\bf 68} (2005) 1405
  [Yad.\ Fiz.\  {\bf 68} (2005) 1461].

\bibitem{Jos}
J.~A.~Vermaseren,
arXiv:math-ph/0010025.

\bibitem{Pivovarov}
A.~A.~Penin and A.~A.~Pivovarov,
Phys.\ Atom.\ Nucl.\  {\bf 64} (2001) 275
[Yad.\ Fiz.\  {\bf 64} (2001) 323];\\
A.~A.~Penin and A.~A.~Pivovarov,
Nucl.\ Phys.\ B {\bf 550} (1999) 375;\\
K.~Melnikov and A.~Yelkhovsky,
Nucl.\ Phys.\ B {\bf 528} (1998) 59.

\bibitem{Hoang:1998uv}
A.~H.~Hoang,
Phys.\ Rev.\ D {\bf 59} (1999) 014039;\\
A.~H.~Hoang and T.~Teubner,
Phys.\ Rev.\ D {\bf 58} (1998) 114023;\\
A.~H.~Hoang,
Phys.\ Rev.\ D {\bf 56} (1997) 5851.

\end{thebibliography}
\end{document}